\documentclass[twocolumn]{aastex701}

\usepackage{threeparttable}
\usepackage{verbatim}  
\usepackage{graphicx}                    
\usepackage{color}                       
\usepackage{breakurl}                         
\usepackage{rotating}
\usepackage{hyperref}
\usepackage{siunitx}

\usepackage{ltablex} 
\usepackage{threeparttablex} 

\newcommand{\lrsp}{Living Rev. Sol. Phys.}

\begin{document}

\title{Detection and Statistics of large-scale Sector Boundary Crossings in the Solar Wind:\\The SBCross Catalog}

\author[orcid=0000-0002-2655-2108,sname='SGH']{Stephan~G.~Heinemann}
\affiliation{Department of Physics, University of Helsinki, Finland}
\affiliation{Institute of Physics, University of Graz, Austria}
\email[show]{stephan.heinemann@hmail.at}

\author[orcid=0000-0003-4867-7558, sname='MT']{Manuela~Temmer} 
\affiliation{Institute of Physics, University of Graz, Austria}
\email{manuela.temmer@uni-graz.at}

\author[orcid=0000-0003-2061-2453, sname='MO']{Mathew~Owens} 
\affiliation{Department of Meteorology, University of Reading, Reading, UK,}
\email{m.j.owens@reading.ac.uk}

\author[orcid=0000-0002-4489-8073, sname='EK']{Emilia~Kilpua} 
\affiliation{Department of Physics, University of Helsinki, Finland}
\email{emilia.kilpua@helsinki.fi}

\author[orcid=0000-0001-8780-0673, sname='CF']{Charles~Farrugia} 
\affiliation{Institute for the Study of Earth, Oceans, and Space, University of New Hampshire, Durham, NH, USA}
\email{Charlie.Farrugia@unh.edu}

\author[orcid=0000-0002-8065-2847, sname='DP']{Daniel~Price} 
\affiliation{Department of Physics, University of Helsinki, Finland}
\email{daniel.price@helsinki.fi}

\begin{abstract}
Heliospheric sector boundary crossings (SBCs), composed of the Heliospheric Current Sheet (HCS), Heliospheric Plasma Sheet (HPS), and surrounding plasma halo, represent fundamental structures in the solar wind that influence interplanetary dynamics and space weather. We present the \textsc{SBCross} catalog, an automated, publicly available database containing 1,076 SBC events derived from in-situ observations spanning more than 30 years (1995--2026), covering Solar Cycles 23 and 24 alongside the rising and maximum phases of Solar Cycle 25. By combining interplanetary magnetic field (IMF) and 255~eV suprathermal electron pitch angle distributions (PAD) data, our algorithm reliably captures large-scale polarity transitions, with $76\%$ of detected SBCs exhibiting clear HPS signatures ($\beta > 1$, $n_{\mathrm{peak}} > 10\mathrm{\,cm^{-3}}$). Across the catalog, we find an average occurrence rate of $\approx 2.6$ events per Carrington rotation, a median event duration of $\approx 9.0$\,hours, and a median physical thickness of $\approx 7.4 \times 10^6$\,km. Furthermore, approximately $\approx 15\%$ of detected SBCs interact with interplanetary coronal mass ejections (CMEs) on average, with this proportion rising up to $33\%$ during solar maximum. These CME-associated events exhibit systematically longer durations (median of $10.0$\,hours), $54\%$ larger physical thicknesses, and higher average solar wind velocities. Notably, all high-velocity SBC events ($> 600\mathrm{\,km\,s^{-1}}$) are exclusively driven by CME interactions, demonstrating local plasma pile-up and structural entanglement resulting from fast CMEs overtaking the ambient slow background. The \textsc{SBCross} catalog serves as an updated successor to the classic SBC list by L. Svalgaard, delivering a standardized, open-access resource for the heliophysics community to systematically investigate sector boundary transitions.
\end{abstract}



\section{Introduction} \label{s:intro}
The heliosphere is continuously being replenished by the solar wind, a magnetized plasma stream expanding outward from the solar corona into interplanetary space \citep{Parker1958}. Embedded within this global outflow is the Heliospheric Current Sheet \citep[HCS; e.g.,][]{Svalgaard1976, Wilcox1980, 2001Smith_HCS}, which represents the fundamental boundary separating regions of opposite magnetic polarity. Enclosing the HCS is the Heliospheric Plasma Sheet \citep[HPS;][]{Burlaga1990}, a region characterized by enhanced proton density, elevated plasma $\beta$, a depressed helium-to-hydrogen ratio ($\mathrm{He/H}$), and reduced magnetic field strength and temperature \citep{Gosling1981, Borrini1981, Winterhalter1994}. Observational manifestations of transitioning across these large-scale structures in \textit{in-situ} data are defined as sector boundary crossings (SBCs). Crucially, SBCs are identified by directional reversals in the interplanetary magnetic field (IMF) accompanied by abrupt shifts in the suprathermal electron heat flux \citep{Crooker1996, Kahler1996, Crooker2004}. Consequently, pitch angle distributions (PADs) of suprathermal electrons are frequently used to trace these magnetic polarity transitions \citep[e.g.,][]{Crooker2003, Gosling2006}.\\

The location where the suprathermal electron strahl flips alignment from parallel to antiparallel (or \textit{vice versa}) relative to the magnetic field is traditionally termed the "true" sector boundary \citep{KahlerLin1994, KahlerLin1995}. While the true sector boundary is often synonymous with the HCS \citep{Behannon1981}, \citet{Crooker2004_HPS} demonstrated that the two features can become spatially or temporally misaligned due to interchange reconnection. Furthermore, a single sector boundary transition may contain multiple embedded HCS crossings \citep{Crooker1993,Lavraud2020}.\\

The structure of the region where the large-scale IMF configuration flips can be understood as a multi-layer hierarchy. At the core lie one or multiple thin HCSs \citep{Crooker1993}, high-$\beta$ current layers lasting from minutes to hours that maintain dynamic pressure balance with their surroundings. Within this localized view, the characteristic drop in magnetic field magnitude across the crossing is compensated primarily by a sharp rise in plasma density \citep{Winterhalter1994}. Individual HCSs exhibit typical thicknesses ranging from $600\mathrm{\,km}$ to $1.1 \times 10^6\mathrm{\,km}$, with an average scale on the order of $10^4\mathrm{\,km}$ \citep{Winterhalter1994, Lepping1996, Blanco2008}. \\

Surrounding the thin HCS is the broader HPS, whose spatial extent spans from $3.2 \times 10^5\mathrm{\,km}$ to $4.7 \times 10^6\mathrm{\,km}$ \citep{Winterhalter1994, Lepping1996, Simunac2012, Wu2019}. High-resolution analyses around the HPS reveal that this central plasma enhancement is itself nested within a broader, lower-density halo spanning several degrees in heliospheric latitude \citep{Bavassano1997, Lacombe2000}. Superposed epoch studies centered on HCS events demonstrate that these nested structures represent the interplanetary extension of coronal helmet streamers, where dense streamer stalks ($\sim 1^\circ\text{--}2^\circ$ wide) are embedded within wider coronal plasma outflows ($\sim 15^\circ\text{--}20^\circ$) \citep{Borrini1981, Gosling1981, Bavassano1997}.\\

The spatial alignment of the HCS and the surrounding dense HPS plays a critical role in modulating the propagation dynamics and geoeffectiveness of Interplanetary Coronal Mass Ejections (CMEs). When fast, transient CMEs erupt from the solar corona, their path through interplanetary space is fundamentally shaped by these large-scale ambient structures. A fast CME overtaking a slow-moving HPS leads to a dynamic collision; the high-density plasma of the sheet resists the expansion of the CME, leading to a plasma pile-up and strong structural deformation of the CME's shock front \citep{Liu2012, Liewer2011}. This structural distortion and compression can amplify the ambient plasma density and plasma $\beta$ at the interface boundary. Crucially, when the magnetic flux rope of an CME interacts with the HCS/HPS structure, the localized magnetic field is compressed, amplifying any southward magnetic field component ($B_z$) within the interaction region \citep{Liu2012,Temmer2025}. When these heavily compressed, high-density structures interact with Earth's magnetosphere, they deliver a spike in dynamic pressure along with possibly long periods of strong southward $B_z$ \citep{Echer2008}. This combination regularly triggers unexpectedly severe, multi-phase geomagnetic storms that are  more intense than those driven by isolated CMEs traveling through a quiet solar wind \citep[\textit{e.g.,}][]{Temmer2026,Kataoka2015}. Several observational and modeling studies have also shown that CMEs are frequently deflected toward the HCS, which can lead to signification deviations in their expected propagation paths and complicate space weather forecasting \citep[e.g.,][]{Isavnin2014,Kay2015a,Winslowetal2016,wang2020deflection}. The HCS / HPS may also substantially affect the propagation, transport and characteristics of solar energetic particles \citep[SEPs; e.g.,][]{Battarbee2017,Liou2024}\\

In this study, we present a comprehensive statistical analysis of SBCs and provide a catalog spanning over 30 years of observations (1995--2026), covering entirely Solar Cycles 23 and 24, as well as the rising and maximum phases of Solar Cycle 25. The study details the distinctive physical properties of SBCs and the surrounding plasma environment derived from a systematic combination of IMF and PAD data. A major focus is the identification and analysis of SBC events that directly interact with interplanetary CMEs. As discussed above, the plasma environment during SBCs can distort the local plasma topology, causing significant variations in CME geoeffectiveness, altering shock propagation dynamics, and creating substantial statistical scatter in their downstream profiles compared to isolated, non-interacting events \citep{Richardson2018, Temmer2026}. By linking our SBC events to an existing in-situ CME catalog, this study offers a reliable statistical baseline for exploring how these large-scale background structures influence, shape, and accelerate severe space weather in the near-Earth environment.

\section{Methodology}\label{s:methods}

\subsection{Data}\label{subs:data}
In this study we utilize low-res OMNI solar wind data \citep[1 hour resolution;][]{papitashvili2020omni} and Electron Pitch Angle Distribution (PAD) data from the Three-Dimensional Plasma instrument \citep[3DP;][]{Lin1995_Wind3DP} on-board the \textit{Wind} spacecraft. Specifically, we are employing the EESA-Low (Electron Electrostatic Analyzer-Low) sensor data to retrieve PAD data in the 255 eV channel. We utilize the full available data from 1995 to 2026.\\

To identify potential associations between the detected SBC events and CMEs, we utilize the Richardson-Cane CME list \citep[R\&C list;][]{2010richardson_RC-list} to extract the disturbance onset and the end of the magnetic obstacle until September 2025.

\subsection{SBC detection}\label{subs:sbcdetection}

To extract and characterize heliospheric SBCs and the associated plasma environments such as the HCS, HPS and surrounding halo, we utilize two independent solar wind data products. The events derived from each product are subsequently combined to construct the final catalog. \\

Primarily, we utilize normalized pitch angle distribution (PAD) data at 255\,eV to derive an asymmetry index ($I_{\mathrm{asym}}^{255\mathrm{eV}}$) from the $15^{\circ}$ (parallel) and $165^{\circ}$ (anti-parallel) bins, defined as:
\begin{equation}
    I_{\mathrm{asym}}^{255\mathrm{eV}} = \frac{\mathrm{PAD}_{15^{\circ}}-\mathrm{PAD}_{165^{\circ}}}{\mathrm{PAD}_{15^{\circ}}+\mathrm{PAD}_{165^{\circ}}}.
\end{equation}
This asymmetry index, which reflects the tentative outward or inward direction of the magnetic field, is binned to a 1-hour resolution to match the solar wind data and is further smoothed using a Gaussian kernel with a sigma ($\sigma$) of 6\,hours. Utilizing this index, we define SBC events as transitions from $0.2$ to $-0.2$ (and vice versa) occurring between two nearly uninterrupted 40-hour periods dominated by a single field direction ($|I_{\mathrm{asym}}^{255\mathrm{eV}}| \geq 0.2$). The duration of the event is defined as the time elapsed between these two transitions. \\

The secondary approach for detecting SBC events utilizes the azimuth angle ($Az$) derived from the interplanetary magnetic field (IMF) components in the Geocentric Solar Ecliptic (GSE) coordinate system, computed as:
\begin{equation}
    Az\mathrm{\,[rad]} = \arctan\left(\frac{B_y}{B_x}\right).
\end{equation}
To remove small-scale fluctuations and isolate large-scale sector transitions, we smooth the computed azimuth angle using a Gaussian kernel ($\sigma = 6,$h). Smoothing is performed using circular averaging to prevent phase-wrapping artifacts at the periodic $0^\circ/360^\circ$ boundary. Data gaps shorter than $6\,\mathrm{h}$ were linearly interpolated, whereas larger gaps were assigned \texttt{NaN} values and explicitly excluded from the Gaussian smoothing process. From the distribution of these angles, we identify four distinct regions associated with inward ($\pi$ to $\pi/2$) and outward ($0$ to $-\pi/2$) fields, as well as their respective transitions ($\pi/2$ to $0$ and $-\pi/2$ to $-\pi$). Similar to the PAD asymmetry index approach, sector boundary transitions are identified as the transition from an outward to an inward field (or vice versa). The periods of outward and inward fields must last nearly uninterruptedly ($>95\%$ of datapoints) for at least 40\,hours. The duration of these transitions is defined as the time interval between the stable outward and inward field periods.\\

The parameters of the detection algorithm were determined through empirical optimization. The smoothing parameter ($\sigma = 6$\,h) removes high-frequency fluctuations while preserving distinct sector boundary signatures. The $40$\,hour threshold was selected as a representative duration for a stable interplanetary magnetic sector ($\sim 1.5-2$ days), preventing short-lived polarity reversals or transient magnetic structures (e.g., small flux ropes or turbulence) from being misidentified as major sector boundary crossings. These choices were optimized by cross-referencing the automated detections against visual inspection and evaluating performance against the historical SBC catalog established by Leif Svalgaard \citep{Svalgaard1972_SBlist} and hosted by the Wilcox Solar Observatory (WSO). Although an SBC catalog exists, a new one is necessary because the Svalgaard list is no longer maintained and provides only approximate boundary crossing times without detailed physical parameters. Sensitivity tests show that slight variations in these thresholds (e.g., $38-42$\,hoursfor polarity duration or $\sigma = 4-8$\,hoursfor smoothing) yield nearly identical event selections, demonstrating that the final catalog is robust against minor parameter choices.

\begin{figure*}[h]
   \centering
   \includegraphics[width=1\linewidth]{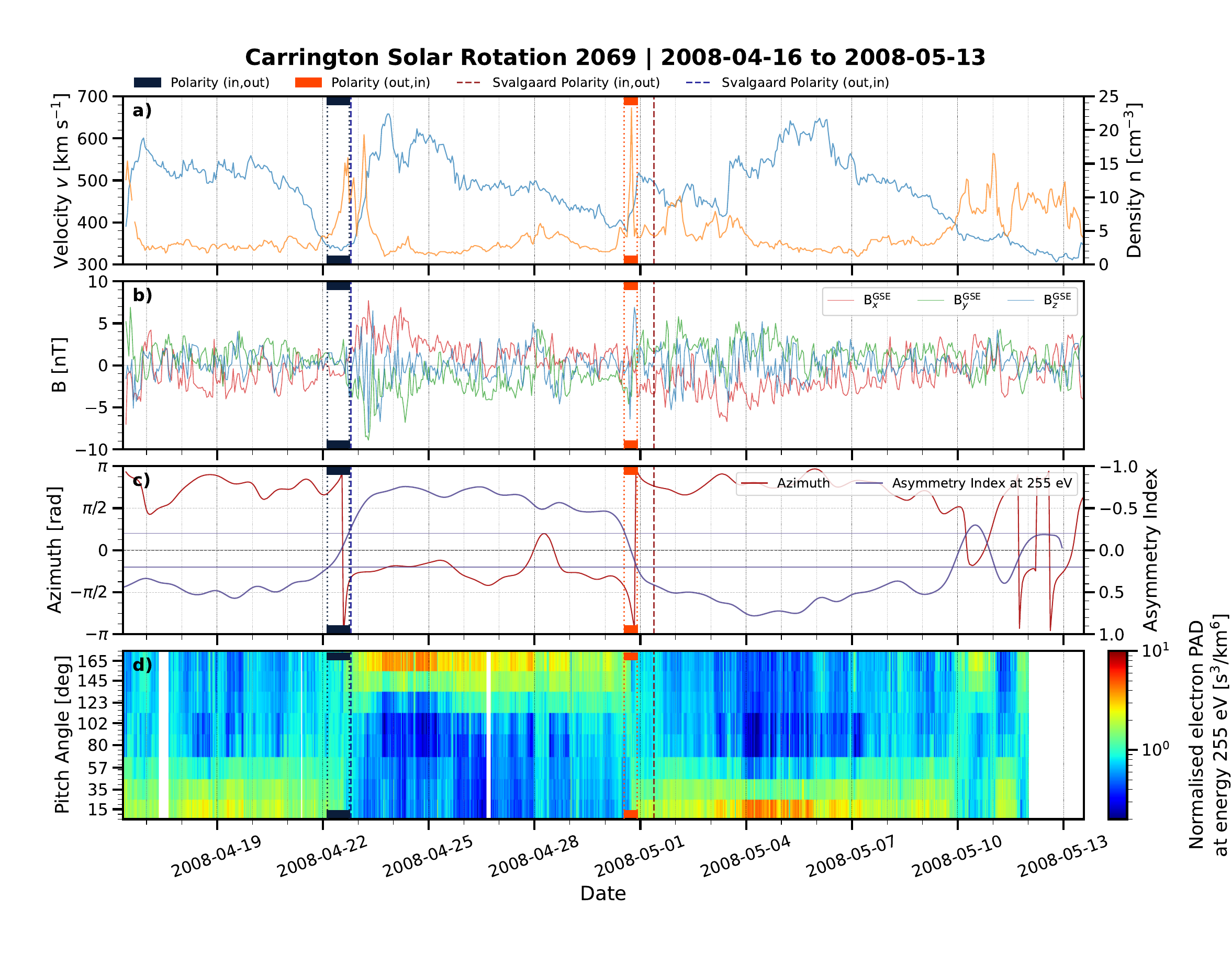}
   \caption{Overview of solar wind plasma and interplanetary magnetic field parameters observed during \textit{Carrington} rotation 2069 (April 16 to May 13, 2008). From top to bottom, the panels display: (a) proton bulk velocity (blue, left axis) and proton number density (orange, right axis); (b) IMF components ($B_x$, $B_y$, $B_z$) in Geocentric Solar Ecliptic (GSE) coordinates; (c) IMF azimuthal angle (dark red, left axis) superimposed with the suprathermal electron asymmetry index evaluated at 255 eV (purple, right axis); and (d) the corresponding normalized electron pitch angle distribution (PAD) spectrogram at 255 eV, color-coded on a logarithmic scale. The solid bars with the dotted lines show the detected SBC intervals and the dashed lines the sector boundary transitions from the L. Svalgaard catalog.}
              \label{fig:overview}%
\end{figure*}

By performing this automated analysis on the available data from the start of 1995 to the end of 2025, we derive two distinct sets of events that are subsequently combined, with confidence based on their detection source. If a transition is simultaneously detected by both the PAD asymmetry index and the IMF azimuth angle, the event is assigned the highest reliability flag ($R=high$), and the duration derived from the asymmetry index is utilized to define the length of the SBC event. When an event is uniquely identified in the PAD asymmetry data product but absent in the IMF azimuth angle data, it is included in the combined catalog with a moderate reliability flag ($R=moderate$). Conversely, if an event is detected solely via the IMF magnetic field data, it is recorded and flagged with the lowest reliability ($R=low$). These reliability flags simply indicate the data source used for detection, allowing community users to filter the SBCross catalog according to their specific research needs. Further, a manual inspection is performed to identify and remove the automated detection artifacts. For all categorized events, their respective physical properties are systematically recorded in the Sector Boundary Crossing (SBCross) catalog. As a representative example of this detection process, Carrington rotation 2069 is illustrated in Figure~\ref{fig:overview}.

\subsection{Computing SBC thickness}
To determine the physical thickness of SBCs, we calculated its cross-sectional width by tracking the dynamic orientation of the solar wind. Because traditional Minimum Variance Analysis can degrade when applied to low-cadence, hourly averaged magnetic data across multi-hour intervals \citep{Sonnerup1998ISSI_MVA}, we adopted a robust geometric approach based on the Parker spiral configuration.Rather than assuming a fixed $45^\circ$ Parker spiral angle, we determined the local Parker spiral angle ($\psi$) which represents the expected orientation of the magnetic sector front relative to the radial direction, using the mean solar wind speed ($V$) measured during the crossing:

\begin{equation}
\psi = \arctan\left(\frac{\Omega_\odot r}{V}\right)
\end{equation}
where $\Omega_\odot \approx 2.865 \times 10^{-6}\text{ rad s}^{-1}$ is the sidereal solar rotation rate and $r \approx 1\text{ AU}$ is the heliocentric distance.

We then projected the velocity along the boundary normal to find the normal impact velocity, $V_{\text{normal}} = V \cos \psi$. The total physical thickness ($L$) was calculated by multiplying this normal velocity by the SBC duration ($\Delta t$):
\begin{equation}
L = V \cos \psi \times \Delta t
\end{equation}
Accounting for changing solar wind speeds avoids projection errors and keeps our thickness calculations accurate throughout the solar cycle.

\section{Results}\label{s:res}

\subsection{Statistics} \label{subs:stats}
For our dataset spanning the period from 1995 to 2025, we identify 1,076 SBC events. Among these, 788 are detected simultaneously in both the PAD and IMF data, 80 are found exclusively in the PAD data, and 208 are identified solely in the IMF data. By filtering these events for properties indicative of clear HPSs \citep[$\beta > 1$ and $n_{\mathrm{peak}} > 10\mathrm{\,cm}^{-3}$; according to][]{Winterhalter1994, Crooker2004_HPS}, we find that 817 ($76\%$) of the detected SBC events seem to include at least one HPS. Of the 817 HPS-associated events, 612 ($75\%$) belong to the high-reliability group, 60 ($7\%$) to the moderate-reliability group, and 145 ($18\%$) to the low-reliability group. \\

\begin{figure}[h]
   \centering
   \includegraphics[width=1\linewidth]{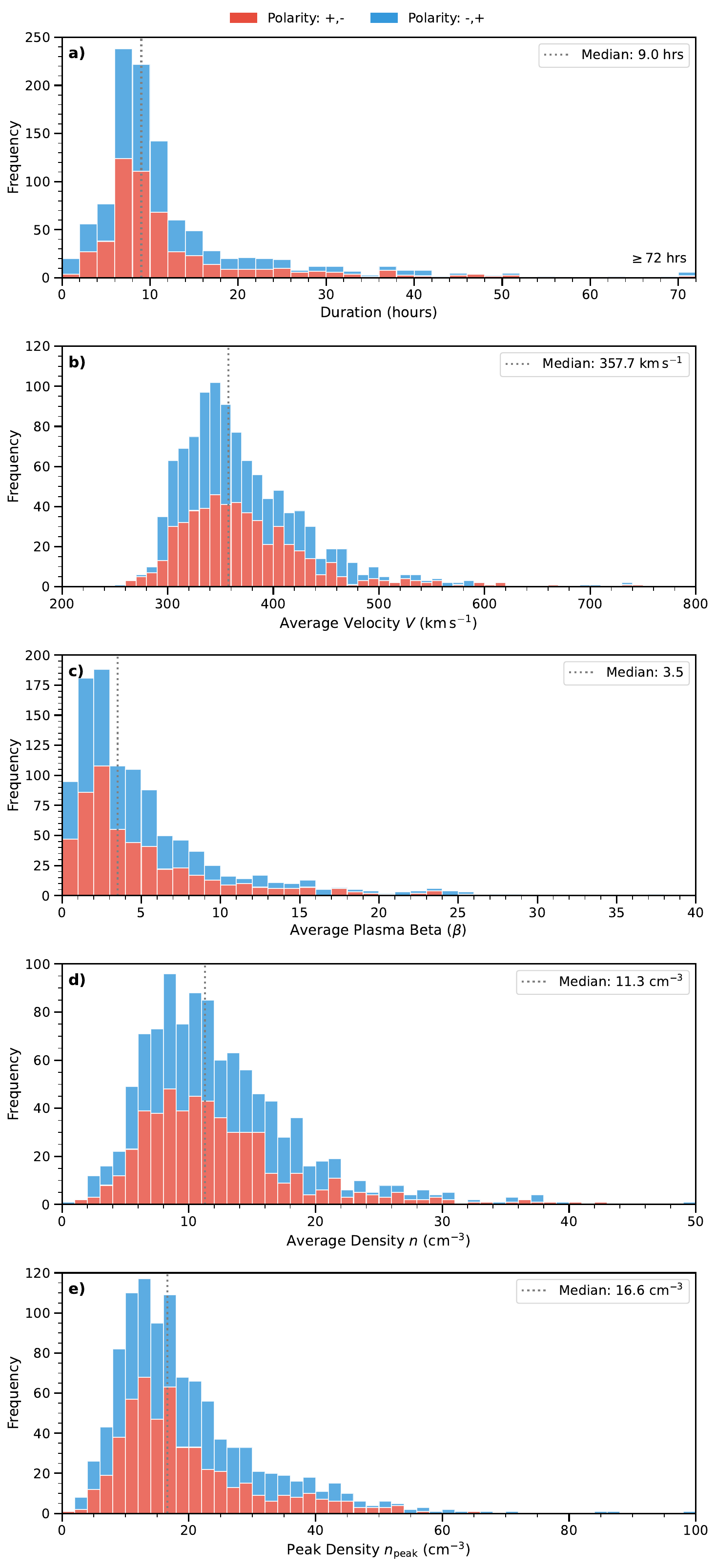}
   \caption{Distribution of SBC crossing characteristics for the 1,076 SBC events, categorized by magnetic polarity orientation ($+,-$ in red; $-,+$ in blue): (a) duration of SBCs, with values exceeding 72~hours grouped into the final bin; (b) average solar wind bulk velocity ($V$); (c) average plasma beta ($\beta$); (d) average proton density ($n$); and (e) peak proton density ($n_{\text{peak}}$). Vertical dashed lines indicate overall median values across all events.}
   
              \label{fig:hist}%
\end{figure}

Figure~\ref{fig:hist} illustrates the distribution of key SBC parameters, specifically the event duration, average solar wind velocity, average plasma beta, average and peak plasma density. The distribution features a median duration of approximately 9\,hours, characterized by a Gaussian-like profile for events lasting less than $\approx24$\,hours, alongside a prominent tail of longer durations. A total of 123 events have a duration exceeding 24 hours and are hereafter referred to as \textit{long-duration SBCs}, with the longest recorded event lasting $128$\,hours($5.3$\,days) in 2004. \\

By averaging the solar wind bulk velocity over each event's duration interval, we find a mean velocity of $\approx370\mathrm{\,km\,s^{-1}}$ and a median velocity of $\approx360\mathrm{\,km\,s^{-1}}$. 95 events show an average plasma beta ($\beta$) below one, suggesting no HPS is contained within. We also find that the detected SBCs are high-density structures with a median average density of $\approx 11\mathrm{\,cm}^{-3}$ and a median peak density of $\approx 17\mathrm{\,cm}^{-3}$. Over the entire baseline, we observe an average occurrence rate of 30 to 40 events per year, which corresponds to roughly $\approx2.6$ events per \textit{Carrington} rotation. \\

\begin{figure}[h]
   \centering
   \includegraphics[width=1\linewidth]{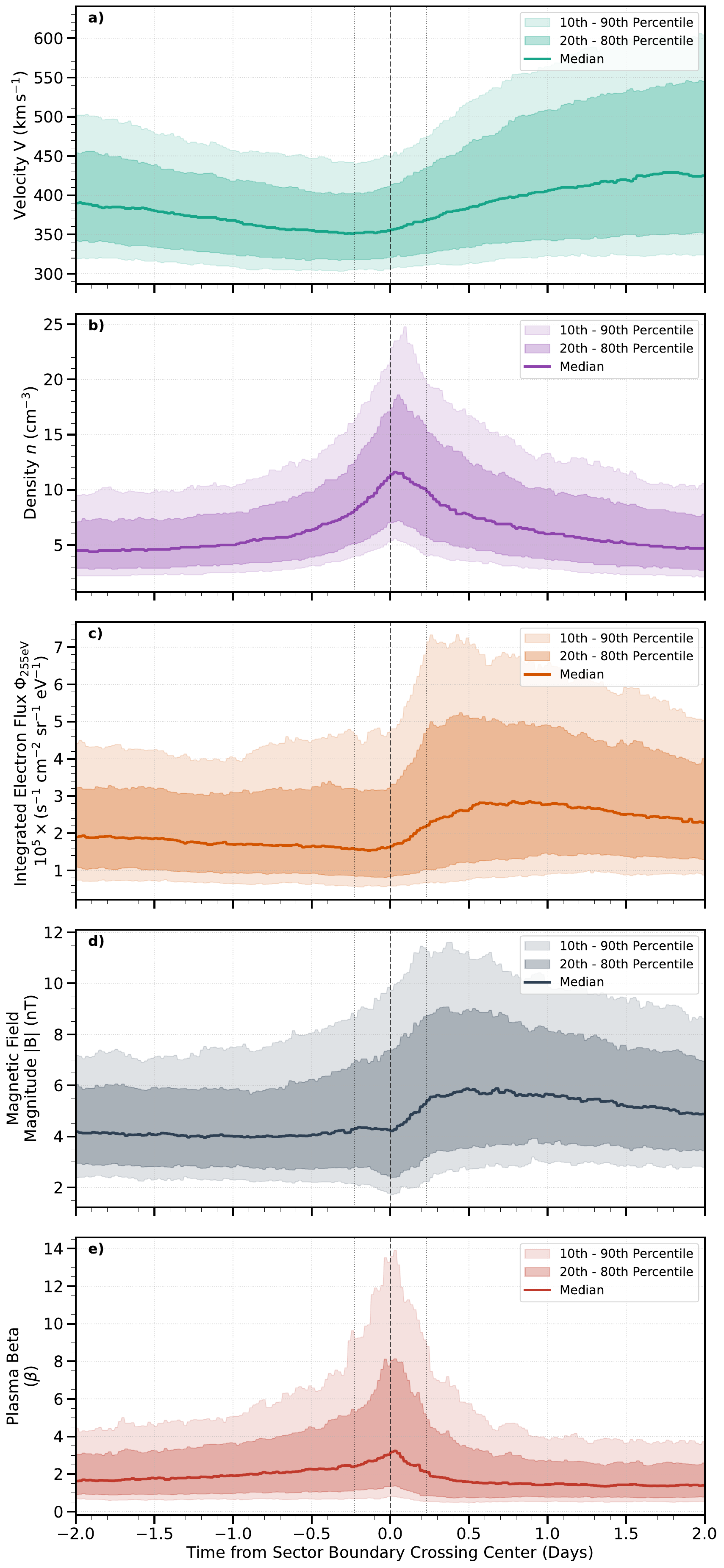}
   \caption{Superposed epoch analysis of solar wind plasma and IMF magnetic field parameters relative to the center of the 1,076 detected SBCs (denoted by the vertical dashed line at Time~$ = 0$ days). The median duration of SBC events is indicated by the vertical dotted lines (at $\pm5.5$ hours). From top to bottom, the panels display the temporal profiles of: (a) solar wind bulk velocity ($V$), (b) solar wind number density ($n$), (c) integrated electron flux in the $255\mathrm{\,eV}$ energy bin, (d) the IMF magnitude $|B|$, and (e) the plasma beta ($\beta$). For all parameters, the solid curves trace the median profiles, while the darker and lighter shaded envelopes represent the $20^{\text{th}}\text{--}80^{\text{th}}$ and $10^{\text{th}}\text{--}90^{\text{th}}$ percentile ranges, respectively. }
              \label{fig:sea}%
\end{figure}

Figure~\ref{fig:sea} displays a superposed epoch analysis (SEA) of the solar wind bulk velocity, number density, integrated electron flux (255 eV channel), magnetic field magnitude and plasma beta for all 1,076 events, centered on the midpoint of the detected SBC intervals. The characteristic minimum/maximum values along with their 10th--90th and 20th--80th percentile ranges are summarized in Table~\ref{tab:sea_summary}. \\

In the velocity profile (Figure~\ref{fig:sea}a), a minimum occurs prior to the center at $\Delta t = -6.92$\,h. Conversely, a pronounced density peak (Figure~\ref{fig:sea}b) follows shortly after the midpoint ($\Delta t = +0.58$\,h), coinciding with the maximum in plasma beta $\beta$ (Figure~\ref{fig:sea}e). The integrated $255\,\mathrm{eV}$ electron flux (Figure~\ref{fig:sea}c) reaches a minimum at $\Delta t = -3.42$\,h. Finally, the magnetic field magnitude $|B|$ (Figure~\ref{fig:sea}d) shows a clear minimum centered on the midpoint ($\Delta t = +0.08$\,h), corresponding to the expected magnetic field drop at the heliospheric current sheet crossing.

\begin{table*}[htbp]
\centering
\caption{Summary of superposed epoch analysis (SEA) extremum values and percentile ranges for all detected SBC events ($N = 1,076$), corresponding to Figure~\ref{fig:sea}. For each parameter, the median minimum or maximum value is listed along with its timing relative to the epoch center ($\Delta t$), as well as the 20th--80th and 10th--90th percentile ranges. Note that the uncertainty in $\Delta t$ due to data cadence is approximately $\pm 1$\,h.}

\label{tab:sea_summary}
\begin{tabular}{llcccc}
\hline\hline
Parameter & Feature & $\Delta t$ & Median & 20th--80th Percentile & 10th--90th Percentile \\
\hline
Bulk Velocity $V$ ($\mathrm{km\,s^{-1}}$) & Min & $-6.92$\,h & 351.0 & [318.0, 402.0] & [304.0, 442.0] \\
Number Density $n$ ($\mathrm{cm^{-3}}$) & Max & $+0.58$\,h & 11.60 & [7.10, 18.20] & [5.60, 23.48] \\
Plasma Beta $\beta$ & Max & $+0.58$\,h & 3.23 & [1.37, 8.14] & [0.80, 13.92] \\
Electron Flux $\Phi_{255\mathrm{eV}}$ ($\mathrm{cm^{-2}\,s^{-1}\,sr^{-1}}$) & Min & $-3.42$\,h & 1.54 & [0.84, 3.15] & [0.60, 4.72] \\
Magnetic Field Magnitude $|B|$ ($\mathrm{nT}$) & Min& $+0.08$\,h & 4.22 & [2.39, 7.39] & [1.71, 9.73] \\
\hline
\end{tabular}
\end{table*}

These distinct multi-instrument signatures show that our methodology successfully detects and extracts SBCs that statistically feature a central HCS often associated with a HPS structure. Furthermore, the velocity profile exhibits an increase after the average SBC interval, which is clearly associated with the onset of a Stream Interaction Region (SIR) and a trailing high-speed solar wind streams (HSSs).  The corresponding post-interval enhancements in $|\mathrm{B}|$ and electron flux appear to map to the compressed plasma of the stream interaction regions and their respective stream interfaces preceding the HSSs \citep[as shown by][]{Crooker2010}. However, distinguishing the density increase caused by the SIR from that of the SBC plasma environment remains difficult. \\

\subsection{SBCs and CMEs} \label{subs:cmes}

Interaction of CMEs with SBCs can significantly alter their geoeffectiveness. In space weather forecasting, particularly the heliospheric current sheet as discussed by \cite{Temmer2025,Temmer2026}, warrants further attention as crucial parameter. As such, we investigate SBC events that can be associated with in-situ measured CMEs as listed in the R\&C list. We define an SBC event as CME-associated if any portion of a CME interval overlaps with the SBC interval, or if the start or end of a CME occurs within 12 hours of the SBC boundary boundaries. We further sort these interaction into three groups, where a CME precedes a SBC (occurs before but has no overlap), a CME succeeds a SBC (occurs after without an overlap) and events where the R\&C listed interval overlaps with the SBC interval. \\

\begin{figure}[h]
   \centering
   \includegraphics[width=1\linewidth]{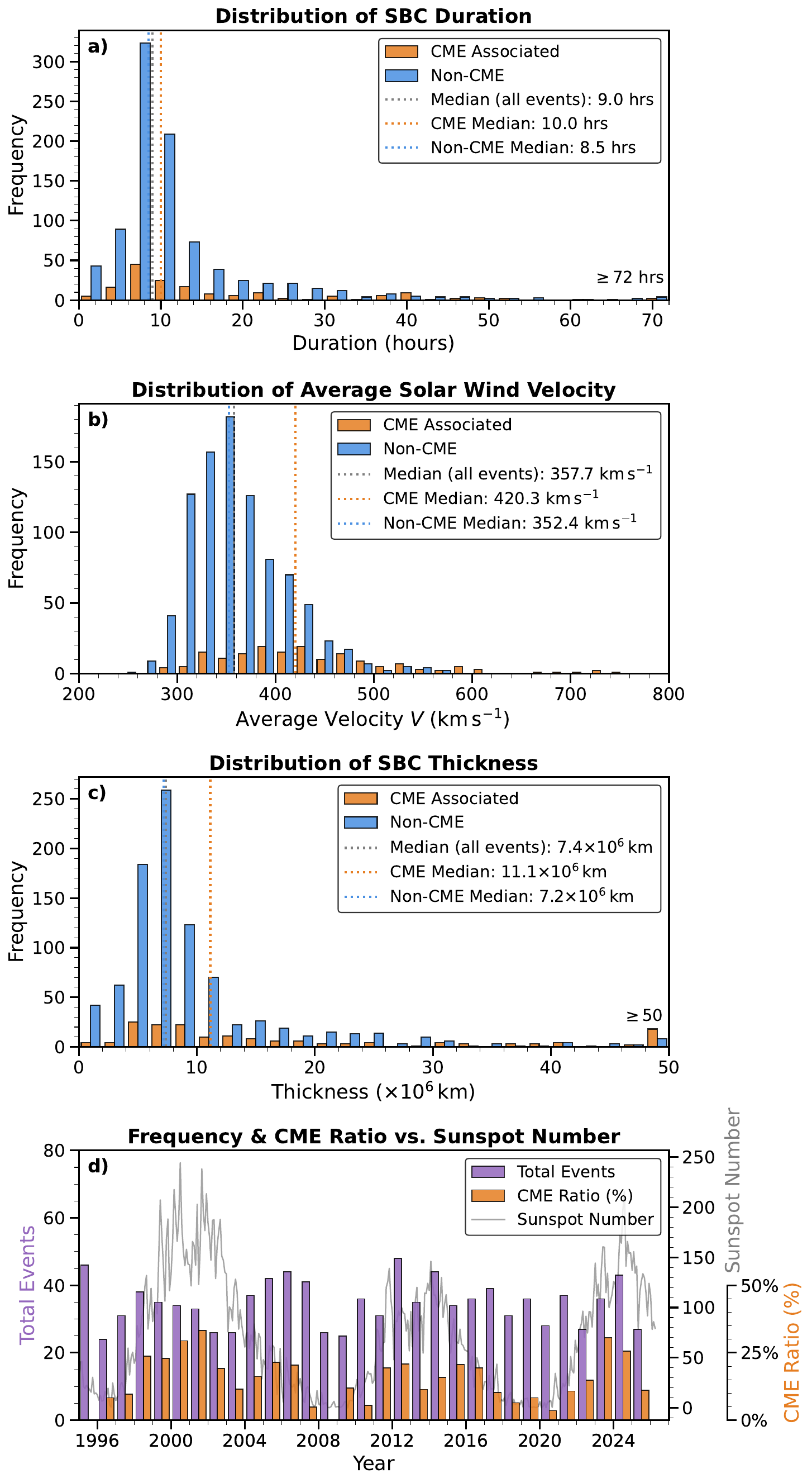}
   \caption{Distribution of SBC characteristics for CME-associated (orange) and non-CME (blue) events: (a) SBC duration, with values exceeding 72~hours grouped into the final bin; (b) average solar wind bulk velocity ($V$); (c) physical thickness of the transition regions, with values exceeding $50\times10^{6}$~km grouped into the final bin; and (d) full time series from 1996 to 2026 comparing the annual frequency of total detected SBC events (blue bars, left axis), the annual CME association ratio (orange dashed line, right axis), and the monthly sunspot number from SILSO\footnote{WDC–SILSO, Royal Observatory of Belgium, International Sunspot Number (SN), Version 2, DOI: 10.24414/qnza-ac80 (CC-BY-NC).} (grey curve, far-right axis). Dotted vertical lines in panels (a)–(c) indicate the overall, CME-associated, and non-CME median values.}
              \label{fig:hist_cme}%
\end{figure}

We find that 166 of the SBC events are interacting with at least one CME (note that we did not check for association with multiple CMEs), which corresponds to about $15\%$ of all recorded events. As expected, the occurrence of CME-associated events is strongly modulated by the solar cycle (Figure~\ref{fig:hist_cme}d). While the annual count of total SBC events shows no solar cycle dependence (Spearman correlation coefficient $\rho = 0.07,~p = 0.699$), CME-associated SBC counts exhibit a strong positive correlation with the annual average sunspot number ($\rho = 0.70,~p < 0.001$). Consequently, the relative fraction of CME-associated SBC events is also strongly correlated with the sunspot number ($\rho = 0.75,~p < 0.001$), reaching up to $33\%$ during solar maximum. Figure~\ref{fig:hist_cme} further shows the difference in duration (a) and solar wind velocity (b) between SBC events with and without associated CMEs. The distribution clearly reveals that CME related events last longer (median duration is $\approx1.5-2$ hours longer) and have a faster average solar wind velocity (about $\approx60\mathrm{\,km\,s^{-1}}$ faster). We further find that the proportion of SBC events associated with CMEs is larger for long duration events ($29\%$) than of events with a duration of below one day ($14\%$). Figure~\ref{fig:hist_cme}c shows the distribution of SBC thickness across all 1,076 events, yielding an overall median thickness of $\approx 7.4 \times 10^6$\,km (10th--90th percentile range: $[3.9, 23.0] \times 10^6$\,km). When categorized by CME association, CME-associated events exhibit a median thickness of $11.1 \times 10^6$\,km, which is roughly $54\%$ larger than that of non-CME events ($7.2 \times 10^6$\,km). Similar to our findings for event duration, we find no direct correlation between the solar cycle (as approximated by the sunspot number) and SBC thickness (not shown). However, solar maximum features a significantly larger proportion of CME-associated events, which are statistically both longer in duration and physically thicker. All results discussed in this section are statistically significant based on Mann-Whitney $U$ tests \citep{Mann1947}, with non-normality confirmed via Shapiro--Wilk tests \citep{Shapiro1965}. A detailed summary of these statistical results is provided in Appendix Table~\ref{tab:event_comparison}.\\

Figure~\ref{fig:scatter_vdur} shows the mean bulk velocity of the SBC events plotted against their duration, with color indicating CME association.We find that bulk velocity and event duration are essentially uncorrelated across all categories ($\rho \approx -0.19$ to $0.09$, with $p > 0.10$ for all CME-associated subsets). However, 17 out of the 22 SBC events with an average velocity exceeding $550\mathrm{\,km\,s^{-1}}$, as well as all events exceeding $600\mathrm{\,km\,s^{-1}}$, overlap with CMEs. Furthermore, CMEs occurring even shortly after an SBC event do not systematically alter the observed average velocity or duration of the SBC structure compared to isolated events. In contrast, SBC events that directly overlap with CMEs exhibit a significantly larger scatter in both duration and velocity, despite lacking any correlation.


\begin{figure}[h]
   \centering
   \includegraphics[width=1\linewidth]{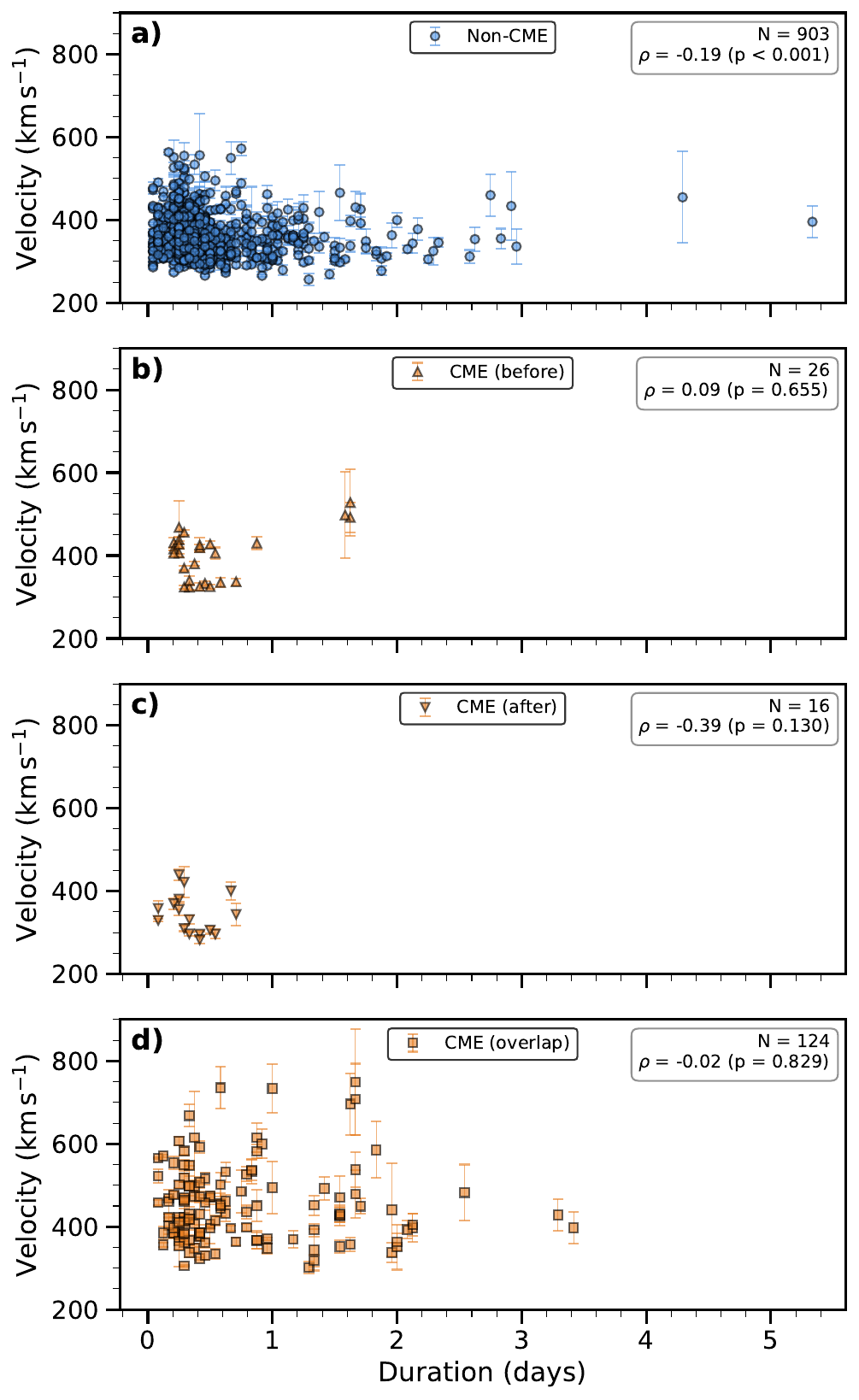}
    \caption{Scatter plots of average solar wind velocity versus SBC duration, categorized by CME association: (a) non-CME events ($N=903$); (b) CME before the SBC ($N=26$); (c) CME after the SBC ($N=16$); and (d) CME overlapping with the SBC ($N=124$). Error bars represent standard deviations.}
              \label{fig:scatter_vdur}%
\end{figure}

\subsection{The SBCross Catalog} \label{subs:catalog}

Using the automated method described in this study, we developed a new, comprehensive catalog of SBCs, including their surrounding plasma properties. The catalog is published online at \url{https://sbcross.helsinki.fi}, hosted by the University of Helsinki, Finland, in collaboration with the University of Graz, Austria. The detection algorithm is permanently archived on Zenodo\footnote{\dataset[10.5281/zenodo.23061201]{https://doi.org/10.5281/zenodo.23061201}}
and maintained on GitHub\footnote{\url{https://github.com/sgheinemann/SBCross}}.Currently, the catalog spans data from 1995 to 2026 and contains the date, duration, and multiple plasma and magnetic field properties, as well as a confidence estimation and CME association. We aim to maintain the catalog through ongoing updates by S.G. Heinemann, D. Price, and M. Temmer. The static dataset is archived on Zenodo\footnote{\dataset[10.5281/zenodo.22078280]{https://doi.org/10.5281/zenodo.22078280}}.

\section{Discussion}\label{s:disc}

Using PAD and IMF data, we identified 1,076 magnetic field sector changes of which $76\%$ are associated with high-density HPS structures. The identified events are compiled in the SBcross catalog. We evaluated the performance of our identification method by benchmarking our catalog against the comprehensive sector boundary list compiled by L. Svalgaard \citep[][]{Svalgaard1972_SBlist}. Out of the 1,076 total events in our catalog, 1,030 fall within the temporal overlap period with Svalgaard’s list (1995 to June 2024, marking its final update), compared to the 844 events recorded by Svalgaard. Overall, 748 entries in the SBCross catalog correspond directly to events in Svalgaard's list, representing an 89\% agreement within a $\pm1.5$-day window of the cataloged timestamps. Additionally, our automated method identified 182 events not present in Svalgaard's list, whereas 96 entries from Svalgaard's catalog were not detected by us. A detailed comparison and visual inspection of these discrepancies revealed two main factors driving the variations between the catalogs. First, events listed by Svalgaard but missed by SBCross often occur during complex intervals with overlapping CMEs or CME-like structures. These disturbances obscure clear sector boundary transitions in PAD or IMF orientation, precluding automated detection, while a small subset showed no distinct SBC signatures at all. Second, SBCross identifies additional boundary crossings because it requires a shorter minimum sector duration before and after a reversal (40 hours compared to Svalgaard's 4 days). This shorter stability requirement allows the algorithm to detect distinct, well-defined SBCs bounded by shorter-lived IMF sectors that Svalgaard's catalog filtered out. Across the full catalog, we find an average of $\approx 2.6$ events per Carrington rotation. This value is within the range reported by WSO (2--4 days), though lower than the $3.4$ events per rotation calculated by \cite{LiouWu2021}, who manually identified 1,197 HCS-focused events from \textit{Wind} data between 1995 and 2020.

From the duration of the SBCs, we estimated their median thickness to be $\approx 7.4 \times 10^6$\,km, with 10th and 90th percentiles of $3.9 \times 10^6$\,km and $2.3 \times 10^7$\,km, respectively. Comparing this to measurements of the HPS, which range from $3.2 \times 10^5$ to $4.7 \times 10^6$\,km \citep{Winterhalter1994, Lepping1996, Simunac2012, Wu2019}, clearly highlights that the intervals defined in this study contain not only a possible HPS but also the surrounding plasma halo. This is further supported by the average plasma beta of $3.5$ within the SBCs. \cite{Klein1980} identified two types of SBCs: those with durations $<10$\,min and those with durations $\ge 3$\,hours. Due to our detection methodology (e.g., smoothed hourly data), all of our detected SBCs fall into the second category.\\

As shown in Figure~\ref{fig:sea}, this approach successfully captures the density enhancement and magnetic field minimum typically associated with the HCS, HPS, and surrounding halo during a sector boundary transition. The superposed epoch analysis yields velocity and density profiles similar to those presented by \citet{Gosling1981} for 23 well-defined sector boundaries.

\citet{Wu2019} noted in their review a clear solar cycle dependence in HPS thickness, characterized by a linear relationship between logarithmic thickness and sunspot number. For the SBCs identified in this study ($76\%$ showing HPS signatures), we find virtually no correlation between sunspot number and either event duration ($\rho \approx  -0.12,~p < 0.001$) or physical thickness ($\rho \approx -0.03,~p = 0.234$).

We further note that SBCs are not exclusively tied to stream interaction regions \citep[SIR;][]{Gosling1981, Borrini1981}. We observe a distinct rise in solar wind velocity approximately one day after the density peak, which is accompanied by a peak in the suprathermal electron flux indicative of a stream interaction region \citep[e.g.,][]{gosling96, Richardson2018}. Subsequent manual inspection reveals that the density peak in SBCs often precedes an SIR, regularly coincides with one, or exhibits a very narrow separation, manifesting as a distinct double-peak structure in the density profile \citep[see also][Figure~3.16]{Schwenn1990}.\\

The solar cycle modulation observed in CME-associated SBC events peaks at approximately $33\%$ during solar maximum and drops nearly to zero during solar minimum, agreeing well with the established solar cycle dependence of CME occurrence rate \citep{Kilpua2017a}. High solar wind velocities ($> 600\mathrm{\,km\,s^{-1}}$) are observed exclusively during direct overlaps of CMEs and SBCs, providing strong in-situ evidence of faster CMEs actively interacting with a slower background SBC. This dynamic collision can cause local plasma pile-up, which expands the event duration and generates highly complex ejecta \citep{Crooker2004_HPS, Lugazetal2017}. 

Interestingly, we find no significant correlation when a CME merely precedes or follows an SBC event without overlapping. This implies that structural modifications require direct physical interaction, supporting the hypothesis that an encroaching CME does not simply displace the sector boundary structure, but embeds itself within or may even replace the local HCS/HPS structure \citep{Forsyth1997}. This mechanism explains the large scatter in duration and velocity observed during SBC events that interact with CMEs; individual CME--SBC collisions are inherently disruptive \citep{Romeo2023,LiuYD2024} and unpredictable, leaving behind a severely distorted or entangled solar wind structure \citep[e.g.,][]{Temmer2021}. 

Furthermore, \citet{Crooker1998} argued that CMEs associated with SBCs may utilize coronal streamer structures as funnels into the heliosphere, a phenomenon that may be directly linked to their enhanced geoeffectiveness.

\section{Summary and Conclusions}\label{s:sum}

In this study, we presented a comprehensive catalog of 1,076 sector boundary crossings (SBCs) spanning from 1995 to 2026. Beyond identifying these distinct events, we characterized their physical properties—focusing on event durations and the surrounding plasma environment, and investigated potential interactions between SBC structures and CMEs. Our primary conclusions are summarized as follows:

\begin{itemize}
    \item We introduced the SBCross catalog, an actively maintained dataset of 1,076 sector boundary crossings published online at \url{https://sbcross.helsinki.fi}. This catalog is designed to serve as a modern, updated successor to the classic sector boundary crossing list compiled by L. Svalgaard.
    \item We find that approximately three out of four SBCs ($76\%$) are associated with at least one HPS structure.
    \item At 1\,AU, SBCs exhibit a median duration of approximately 9\,hours, with a notable population of long-duration events ($>1$~day) accounting for $11\%$ of all cases. Beyond duration, our broader statistical comparison across plasma and magnetic parameters (including velocity, density, plasma beta, and physical thickness) demonstrates significant systematic differences between event sub-populations.
    
    \item On average, approximately $15\%$ of the identified SBC events were directly associated with CMEs, with this fraction reaching up to $33\%$ during solar maximum. These interaction events are characterized by systematically longer durations and higher average velocities compared to isolated SBC crossings. Notably, all high-velocity SBC events ($>600\mathrm{\,km\,s^{-1}}$) were found to be exclusively driven by these CME interactions.
\end{itemize}

\begin{acknowledgments}
 We acknowledge use of NASA/GSFC's Space Physics Data Facility's OMNIWeb/CDAWeb service, and OMNI data. SGH acknowledges funding from the Research Council of Finland (Academy Fellowship) [370747; RIB-Wind]. SGH thanks his chair for always having his back. This research was funded in whole or in part by the Austrian Science Fund (FWF) [10.55776/PAT2280725]. MO is  part-funded by Science and Technology Facilities Council (STFC) grant number UKRI1207 and Natural Environment Research Council (NERC) grant number NE/Y001052/1.  SGH, EK and DP acknowledge the Research Council of Finland Centre of Excellence project SpaceResilience [1374096]. ERC Synergy project GERACLE (4100617), funded by the EU under grant agreement 101166910. Open access funded by Helsinki University Library.
\end{acknowledgments}

\begin{contribution}

SGH conducted the analysis and led the manuscript preparation. DP prepared the online catalog. MT contributed the core concept for this study. All authors participated in revising and editing the manuscript. 


\end{contribution}

%



\appendix

\section{Additional Figures}
\begin{table}[htbp]
\centering
\small
\setlength{\tabcolsep}{3pt}
\caption{Statistical summary comparing All, CME-associated, and Non-CME events split by non-parametric (Median) and parametric (Mean) metrics. Non-normality (Shapiro-Wilk test $p < 0.001$) is denoted by $\checkmark$ for All/CME/Non-CME groups. IQR represents the Interquartile Range ($75^{\text{th}} - 25^{\text{th}}$ percentile), and $\pm\text{SD}$ denotes the standard deviation. The $p$-values are derived from a two-sided Mann-Whitney $U$ test comparing CME vs. Non-CME events.}
\label{tab:event_comparison}
\begin{tabular}{l r@{\hskip 2pt}c@{\hskip 2pt}l r@{\hskip 2pt}c@{\hskip 2pt}l r@{\hskip 2pt}c@{\hskip 2pt}l c c}
\hline
\textbf{Parameter} & \multicolumn{3}{c}{\textbf{All Events}} & \multicolumn{3}{c}{\textbf{CME Events}} & \multicolumn{3}{c}{\textbf{Non-CME Events}} & \textbf{Shapiro-Wilk} & \textbf{Mann-Whitney $U$} \\
\hline
\multicolumn{12}{l}{\textbf{Medians and Interquartile Ranges [IQR]}} \\
\hline
SBC Duration [h] & $9.00$ &  & $[7.00]$ & $10.00$ &  & $[14.00]$ & $8.50$ &  & $[6.00]$ & $\checkmark$ / $\checkmark$ / $\checkmark$ & $<0.001$ \\
Avg Velocity [$\mathrm{km\,s^{-1}}$] & $357.67$ &  & $[71.38]$ & $420.33$ &  & $[108.91]$ & $352.44$ &  & $[61.93]$ & $\checkmark$ / $\checkmark$ / $\checkmark$ & $<0.001$ \\
Avg Density [$\mathrm{cm^{-3}}$] & $11.26$ &  & $[7.39]$ & $9.15$ &  & $[7.00]$ & $11.58$ &  & $[7.11]$ & $\checkmark$ / $\checkmark$ / $\checkmark$ & $<0.001$ \\
Peak Density [$\mathrm{cm^{-3}}$] & $16.60$ &  & $[12.30]$ & $16.80$ &  & $[13.80]$ & $16.60$ &  & $[12.00]$ & $\checkmark$ / $\checkmark$ / $\checkmark$ & $0.319$ \\
Avg Beta & $3.51$ &  & $[4.48]$ & $1.37$ &  & $[2.10]$ & $4.11$ &  & $[4.80]$ & $\checkmark$ / $\checkmark$ / $\checkmark$ & $<0.001$ \\
SBC Thickness [$10^6\,\mathrm{km}$] & $7.38$ &  & $[5.55]$ & $11.13$ &  & $[17.21]$ & $7.18$ &  & $[4.86]$ & $\checkmark$ / $\checkmark$ / $\checkmark$ & $<0.001$ \\
\hline
\multicolumn{12}{l}{\textbf{Means and Standard Deviations [$\pm$SD]}} \\
\hline
SBC Duration [h] & $12.37$ & $\pm$ & $11.63$ & $16.60$ & $\pm$ & $14.88$ & $11.60$ & $\pm$ & $10.77$ & $\checkmark$ / $\checkmark$ / $\checkmark$ & $<0.001$ \\
Avg Velocity [$\mathrm{km\,s^{-1}}$] & $372.53$ & $\pm$ & $63.61$ & $433.91$ & $\pm$ & $92.83$ & $361.25$ & $\pm$ & $48.92$ & $\checkmark$ / $\checkmark$ / $\checkmark$ & $<0.001$ \\
Avg Density [$\mathrm{cm^{-3}}$] & $12.53$ & $\pm$ & $6.48$ & $10.71$ & $\pm$ & $6.86$ & $12.87$ & $\pm$ & $6.36$ & $\checkmark$ / $\checkmark$ / $\checkmark$ & $<0.001$ \\
Peak Density [$\mathrm{cm^{-3}}$] & $19.76$ & $\pm$ & $11.71$ & $19.90$ & $\pm$ & $13.92$ & $19.73$ & $\pm$ & $11.26$ & $\checkmark$ / $\checkmark$ / $\checkmark$ & $0.319$ \\
Avg Beta & $5.22$ & $\pm$ & $5.35$ & $2.34$ & $\pm$ & $3.00$ & $5.76$ & $\pm$ & $5.52$ & $\checkmark$ / $\checkmark$ / $\checkmark$ & $<0.001$ \\
SBC Thickness [$10^6\,\mathrm{km}$] & $11.31$ & $\pm$ & $12.59$ & $19.51$ & $\pm$ & $19.88$ & $9.80$ & $\pm$ & $10.04$ & $\checkmark$ / $\checkmark$ / $\checkmark$ & $<0.001$ \\
\hline
\end{tabular}
\end{table}

\clearpage



\begin{thebibliography}{}
\expandafter\ifx\csname natexlab\endcsname\relax\def\natexlab#1{#1}\fi
\providecommand{\url}[1]{\href{#1}{#1}}
\providecommand{\dodoi}[1]{doi:~\href{http://doi.org/#1}{\nolinkurl{#1}}}
\providecommand{\doeprint}[1]{\href{http://ascl.net/#1}{\nolinkurl{http://ascl.net/#1}}}
\providecommand{\doarXiv}[1]{\href{https://arxiv.org/abs/#1}{\nolinkurl{https://arxiv.org/abs/#1}}}

\bibitem[{M. {Battarbee} {et~al.}(2017){Battarbee}, {Dalla}, \& {Marsh}}]{Battarbee2017}
{Battarbee}, M., {Dalla}, S., \& {Marsh}, M.~S. 2017, \bibinfo{title}{{Solar Energetic Particle Transport Near a Heliospheric Current Sheet},} \apj, 836, 138, \dodoi{10.3847/1538-4357/836/1/138}

\bibitem[{B. {Bavassano} {et~al.}(1997){Bavassano}, {Woo}, \& {Bruno}}]{Bavassano1997}
{Bavassano}, B., {Woo}, R., \& {Bruno}, R. 1997, \bibinfo{title}{{Heliospheric plasma sheet and coronal streamers},} \grl, 24, 1655, \dodoi{10.1029/97GL01630}

\bibitem[{K.~W. {Behannon} {et~al.}(1981){Behannon}, {Neubauer}, \& {Barnstorf}}]{Behannon1981}
{Behannon}, K.~W., {Neubauer}, F.~M., \& {Barnstorf}, H. 1981, \bibinfo{title}{{Fine-scale characteristics of interplanetary sector boundaries},} \jgr, 86, 3273, \dodoi{10.1029/JA086iA05p03273}

\bibitem[{J.~J. {Blanco} {et~al.}(2008){Blanco}, {Rodriguez-Pacheco}, {Hidalgo}, \& {Sequeiros}}]{Blanco2008}
{Blanco}, J.~J., {Rodriguez-Pacheco}, J., {Hidalgo}, M.~A., \& {Sequeiros}, J. 2008, \bibinfo{title}{{Monitoring the heliospheric current sheet local structure for the years 1995 to 2001},} Journal of Atmospheric and Solar-Terrestrial Physics, 70, 226, \dodoi{10.1016/j.jastp.2007.08.030}

\bibitem[{G. {Borrini} {et~al.}(1981){Borrini}, {Gosling}, {Bame}, {Feldman}, \& {Wilcox}}]{Borrini1981}
{Borrini}, G., {Gosling}, J.~T., {Bame}, S.~J., {Feldman}, W.~C., \& {Wilcox}, J.~M. 1981, \bibinfo{title}{{Solar wind helium and hydrogen structure near the heliospheric current sheet: A signal of coronal streamers at 1 AU},} \jgr, 86, 4565, \dodoi{10.1029/JA086iA06p04565}

\bibitem[{L.~F. {Burlaga} {et~al.}(1990){Burlaga}, {Mish}, \& {Whang}}]{Burlaga1990}
{Burlaga}, L.~F., {Mish}, W.~H., \& {Whang}, Y.~C. 1990, \bibinfo{title}{{Coalescence of recurrent streams of different sizes and amplitudes},} \jgr, 95, 4247, \dodoi{10.1029/JA095iA04p04247}

\bibitem[{N.~U. {Crooker} {et~al.}(2010){Crooker}, {Appleton}, {Schwadron}, \& {Owens}}]{Crooker2010}
{Crooker}, N.~U., {Appleton}, E.~M., {Schwadron}, N.~A., \& {Owens}, M.~J. 2010, \bibinfo{title}{{Suprathermal electron flux peaks at stream interfaces: Signature of solar wind dynamics or tracer for open magnetic flux transport on the Sun?},} Journal of Geophysical Research (Space Physics), 115, A11101, \dodoi{10.1029/2010JA015496}

\bibitem[{N.~U. {Crooker} {et~al.}(1998){Crooker}, {Gosling}, \& {Kahler}}]{Crooker1998}
{Crooker}, N.~U., {Gosling}, J.~T., \& {Kahler}, S.~W. 1998, \bibinfo{title}{{Magnetic clouds at sector boundaries},} \jgr, 103, 301, \dodoi{10.1029/97JA02774}

\bibitem[{N.~U. {Crooker} {et~al.}(2004{\natexlab{a}}){Crooker}, {Huang}, {Lamassa}, {Larson}, {Kahler}, \& {Spence}}]{Crooker2004_HPS}
{Crooker}, N.~U., {Huang}, C.-L., {Lamassa}, S.~M., {et~al.} 2004{\natexlab{a}}, \bibinfo{title}{{Heliospheric plasma sheets},} Journal of Geophysical Research (Space Physics), 109, A03107, \dodoi{10.1029/2003JA010170}

\bibitem[{N.~U. {Crooker} {et~al.}(2004{\natexlab{b}}){Crooker}, {Kahler}, {Larson}, \& {Lin}}]{Crooker2004}
{Crooker}, N.~U., {Kahler}, S.~W., {Larson}, D.~E., \& {Lin}, R.~P. 2004{\natexlab{b}}, \bibinfo{title}{{Large-scale magnetic field inversions at sector boundaries},} Journal of Geophysical Research (Space Physics), 109, A03108, \dodoi{10.1029/2003JA010278}

\bibitem[{N.~U. {Crooker} {et~al.}(2003){Crooker}, {Larson}, {Kahler}, {Lamassa}, \& {Spence}}]{Crooker2003}
{Crooker}, N.~U., {Larson}, D.~E., {Kahler}, S.~W., {Lamassa}, S.~M., \& {Spence}, H.~E. 2003, \bibinfo{title}{{Suprathermal electron isotropy in high-beta solar wind and its role in heat flux dropouts},} \grl, 30, 1619, \dodoi{10.1029/2003GL017036}

\bibitem[{N.~U. {Crooker} {et~al.}(1996){Crooker}, {Lazarus}, {Lepping}, {Ogilvie}, {Steinberg}, {Szabo}, \& {Onsager}}]{Crooker1996}
{Crooker}, N.~U., {Lazarus}, A.~J., {Lepping}, R.~P., {et~al.} 1996, \bibinfo{title}{{A two-stream, four-sector, recurrence pattern: Implications from WIND for the 22-year geomagnetic activity cycle},} \grl, 23, 1275, \dodoi{10.1029/96GL00031}

\bibitem[{N.~U. {Crooker} {et~al.}(1993){Crooker}, {Siscoe}, {Shodhan}, {Webb}, {Gosling}, \& {Smith}}]{Crooker1993}
{Crooker}, N.~U., {Siscoe}, G.~L., {Shodhan}, S., {et~al.} 1993, \bibinfo{title}{{Multiple heliospheric current sheets and coronal streamer belt dynamics},} \jgr, 98, 9371, \dodoi{10.1029/93JA00636}

\bibitem[{E. {Echer} {et~al.}(2008){Echer}, {Gonzalez}, {Tsurutani}, \& {Gonzalez}}]{Echer2008}
{Echer}, E., {Gonzalez}, W.~D., {Tsurutani}, B.~T., \& {Gonzalez}, A.~L.~C. 2008, \bibinfo{title}{{Interplanetary conditions causing intense geomagnetic storms (Dst <= -100 nT) during solar cycle 23 (1996-2006)},} Journal of Geophysical Research (Space Physics), 113, A05221, \dodoi{10.1029/2007JA012744}

\bibitem[{R.~J. {Forsyth} {et~al.}(1997){Forsyth}, {Balogh}, {Smith}, \& {Gosling}}]{Forsyth1997}
{Forsyth}, R.~J., {Balogh}, A., {Smith}, E.~J., \& {Gosling}, J.~T. 1997, \bibinfo{title}{{Ulysses observations of the northward extension of the heliospheric current sheet},} \grl, 24, 3101, \dodoi{10.1029/97GL03099}

\bibitem[{J.~T. {Gosling}(1996){Gosling}}]{gosling96}
{Gosling}, J.~T. 1996, \bibinfo{title}{{Corotating and Transient Solar Wind Flows in Three Dimensions},} \araa, 34, 35, \dodoi{10.1146/annurev.astro.34.1.35}

\bibitem[{J.~T. {Gosling} {et~al.}(1981){Gosling}, {Borrini}, {Asbridge}, {Bame}, {Feldman}, \& {Hansen}}]{Gosling1981}
{Gosling}, J.~T., {Borrini}, G., {Asbridge}, J.~R., {et~al.} 1981, \bibinfo{title}{{Coronal streamers in the solar wind at 1 AU},} \jgr, 86, 5438, \dodoi{10.1029/JA086iA07p05438}

\bibitem[{J.~T. {Gosling} {et~al.}(2006){Gosling}, {Eriksson}, \& {Schwenn}}]{Gosling2006}
{Gosling}, J.~T., {Eriksson}, S., \& {Schwenn}, R. 2006, \bibinfo{title}{{Petschek-type magnetic reconnection exhausts in the solar wind well inside 1 AU: Helios},} Journal of Geophysical Research (Space Physics), 111, A10102, \dodoi{10.1029/2006JA011863}

\bibitem[{A. {Isavnin} {et~al.}(2014){Isavnin}, {Vourlidas}, \& {Kilpua}}]{Isavnin2014}
{Isavnin}, A., {Vourlidas}, A., \& {Kilpua}, E.~K.~J. 2014, \bibinfo{title}{{Three-Dimensional Evolution of Flux-Rope CMEs and Its Relation to the Local Orientation of the Heliospheric Current Sheet},} \solphys, 289, 2141, \dodoi{10.1007/s11207-013-0468-4}

\bibitem[{S. {Kahler} \& R.~P. {Lin}(1994){Kahler} \& {Lin}}]{KahlerLin1994}
{Kahler}, S., \& {Lin}, R.~P. 1994, \bibinfo{title}{{The determination of interplanetary magnetic field polarities around sector boundaries using E > 2 keV electrons},} \grl, 21, 1575, \dodoi{10.1029/94GL01362}

\bibitem[{S.~W. {Kahler} {et~al.}(1996){Kahler}, {Crooker}, \& {Gosling}}]{Kahler1996}
{Kahler}, S.~W., {Crooker}, N.~U., \& {Gosling}, J.~T. 1996, \bibinfo{title}{{The topology of intrasector reversals of the interplanetary magnetic field},} \jgr, 101, 24373, \dodoi{10.1029/96JA02232}

\bibitem[{S.~W. {Kahler} \& R.~P. {Lin}(1995){Kahler} \& {Lin}}]{KahlerLin1995}
{Kahler}, S.~W., \& {Lin}, R.~P. 1995, \bibinfo{title}{{An Examination of Directional Discontinuities and Magnetic Polarity Changes around Interplanetary Sector Boundaries Using E > 2 keV Electrons},} \solphys, 161, 183, \dodoi{10.1007/BF00732092}

\bibitem[{R. {Kataoka} {et~al.}(2015){Kataoka}, {Shiota}, {Kilpua}, \& {Keika}}]{Kataoka2015}
{Kataoka}, R., {Shiota}, D., {Kilpua}, E., \& {Keika}, K. 2015, \bibinfo{title}{{Pileup accident hypothesis of magnetic storm on 17 March 2015},} \grl, 42, 5155, \dodoi{10.1002/2015GL064816}

\bibitem[{C. {Kay} {et~al.}(2015){Kay}, {Opher}, \& {Evans}}]{Kay2015a}
{Kay}, C., {Opher}, M., \& {Evans}, R.~M. 2015, \bibinfo{title}{{Global Trends of CME Deflections Based on CME and Solar Parameters},} \apj, 805, 168, \dodoi{10.1088/0004-637X/805/2/168}

\bibitem[{E. {Kilpua} {et~al.}(2017){Kilpua}, {Koskinen}, \& {Pulkkinen}}]{Kilpua2017a}
{Kilpua}, E., {Koskinen}, H. E.~J., \& {Pulkkinen}, T.~I. 2017, \bibinfo{title}{{Coronal mass ejections and their sheath regions in interplanetary space},} \lrsp, 14, 5, \dodoi{10.1007/s41116-017-0009-6}

\bibitem[{L. {Klein} \& L.~F. {Burlaga}(1980){Klein} \& {Burlaga}}]{Klein1980}
{Klein}, L., \& {Burlaga}, L.~F. 1980, \bibinfo{title}{{Interplanetary sector boundaries 1971-1973},} \jgr, 85, 2269, \dodoi{10.1029/JA085iA05p02269}

\bibitem[{C. {Lacombe} {et~al.}(2000){Lacombe}, {Salem}, {Mangeney}, {Steinberg}, {Maksimovic}, \& {Bosqued}}]{Lacombe2000}
{Lacombe}, C., {Salem}, C., {Mangeney}, A., {et~al.} 2000, \bibinfo{title}{{Latitudinal distribution of the solar wind properties in the low- and high-pressure regimes: Wind observations},} Annales Geophysicae, 18, 852, \dodoi{10.1007/s00585-000-0852-x}

\bibitem[{B. {Lavraud} {et~al.}(2020){Lavraud}, {Fargette}, {R{\'e}ville}, {Szabo}, {Huang}, {Rouillard}, {Viall}, {Phan}, {Kasper}, {Bale}, {Berthomier}, {Bonnell}, {Case}, {Dudok de Wit}, {Eastwood}, {G{\'e}not}, {Goetz}, {Griton}, {Halekas}, {Harvey}, {Kieokaew}, {Klein}, {Korreck}, {Kouloumvakos}, {Larson}, {Lavarra}, {Livi}, {Louarn}, {MacDowall}, {Maksimovic}, {Malaspina}, {Nieves-Chinchilla}, {Pinto}, {Poirier}, {Pulupa}, {Raouafi}, {Stevens}, {Toledo-Redondo}, \& {Whittlesey}}]{Lavraud2020}
{Lavraud}, B., {Fargette}, N., {R{\'e}ville}, V., {et~al.} 2020, \bibinfo{title}{{The Heliospheric Current Sheet and Plasma Sheet during Parker Solar Probe's First Orbit},} \apjl, 894, L19, \dodoi{10.3847/2041-8213/ab8d2d}

\bibitem[{R.~P. {Lepping} {et~al.}(1996){Lepping}, {Szabo}, {Peredo}, \& {Hoeksema}}]{Lepping1996}
{Lepping}, R.~P., {Szabo}, A., {Peredo}, M., \& {Hoeksema}, J.~T. 1996, \bibinfo{title}{{Large-scale properties and solar connection of the heliospheric current and plasma sheets: WIND observations},} \grl, 23, 1199, \dodoi{10.1029/96GL00658}

\bibitem[{P.~C. {Liewer} {et~al.}(2011){Liewer}, {Hall}, {Howard}, {De Jong}, {Thompson}, \& {Thernisien}}]{Liewer2011}
{Liewer}, P.~C., {Hall}, J.~R., {Howard}, R.~A., {et~al.} 2011, \bibinfo{title}{{Stereoscopic analysis of STEREO/SECCHI data for CME trajectory determination},} Journal of Atmospheric and Solar-Terrestrial Physics, 73, 1173, \dodoi{10.1016/j.jastp.2010.09.004}

\bibitem[{R.~P. {Lin} {et~al.}(1995){Lin}, {Anderson}, {Ashford}, {Carlson}, {Curtis}, {Ergun}, {Larson}, {McFadden}, {McCarthy}, {Parks}, {R{\`e}me}, {Bosqued}, {Coutelier}, {Cotin}, {D'Uston}, {Wenzel}, {Sanderson}, {Henrion}, {Ronnet}, \& {Paschmann}}]{Lin1995_Wind3DP}
{Lin}, R.~P., {Anderson}, K.~A., {Ashford}, S., {et~al.} 1995, \bibinfo{title}{{A Three-Dimensional Plasma and Energetic Particle Investigation for the Wind Spacecraft},} \ssr, 71, 125, \dodoi{10.1007/BF00751328}

\bibitem[{K. {Liou} \& C.-C. {Wu}(2021){Liou} \& {Wu}}]{LiouWu2021}
{Liou}, K., \& {Wu}, C.-C. 2021, \bibinfo{title}{{Characteristics of the Heliospheric Current Sheets at the Sector Boundaries: Wind Observations from 1995-2020},} \apj, 920, 39, \dodoi{10.3847/1538-4357/ac1586}

\bibitem[{K. {Liou} \& C.-C. {Wu}(2024){Liou} \& {Wu}}]{Liou2024}
{Liou}, K., \& {Wu}, C.-C. 2024, \bibinfo{title}{{Solar Energetic Particle and the Heliospheric Current Sheet},} \apj, 966, 16, \dodoi{10.3847/1538-4357/ad33c2}

\bibitem[{Y.~D. Liu {et~al.}(2012)Liu, Luhmann, M\"{o}stl, Martinez-Oliveros, Bale, Lin, Harrison, Temmer, Webb, \& Odstr{\v{c}}il}]{Liu2012}
Liu, Y.~D., Luhmann, J.~G., M\"{o}stl, C., {et~al.} 2012, \bibinfo{title}{{INTERACTIONS} {BETWEEN} {CORONAL} {MASS} {EJECTIONS} {VIEWED} {IN} {COORDINATED} {IMAGING} {AND} {IN} {SITU} {OBSERVATIONS},} The Astrophysical Journal, 746, L15, \dodoi{10.1088/2041-8205/746/2/l15}

\bibitem[{Y.~D. {Liu} {et~al.}(2024){Liu}, {Zhu}, {Ran}, {Hu}, {Liu}, {Zhao}, {Wang}, {Stevens}, \& {Bale}}]{LiuYD2024}
{Liu}, Y.~D., {Zhu}, B., {Ran}, H., {et~al.} 2024, \bibinfo{title}{{Direct In Situ Measurements of a Fast Coronal Mass Ejection and Associated Structures in the Corona},} \apj, 963, 85, \dodoi{10.3847/1538-4357/ad1e56}

\bibitem[{N. {Lugaz} {et~al.}(2017){Lugaz}, {Temmer}, {Wang}, \& {Farrugia}}]{Lugazetal2017}
{Lugaz}, N., {Temmer}, M., {Wang}, Y., \& {Farrugia}, C.~J. 2017, \bibinfo{title}{{The Interaction of Successive Coronal Mass Ejections: A Review},} \solphys, 292, 64, \dodoi{10.1007/s11207-017-1091-6}

\bibitem[{H.~B. Mann \& D.~R. Whitney(1947)Mann \& Whitney}]{Mann1947}
Mann, H.~B., \& Whitney, D.~R. 1947, \bibinfo{title}{On a Test of Whether one of Two Random Variables is Stochastically Larger than the Other,} The Annals of Mathematical Statistics, 18, 50, \dodoi{10.1214/aoms/1177730491}

\bibitem[{N.~E. Papitashvili \& J.~H. King(2020)Papitashvili \& King}]{papitashvili2020omni}
Papitashvili, N.~E., \& King, J.~H. 2020, {OMNI Hourly Data [1996-2026]},, \url{https://doi.org/10.48322/1shr-ht18} NASA Space Physics Data Facility

\bibitem[{E.~N. {Parker}(1958){Parker}}]{Parker1958}
{Parker}, E.~N. 1958, \bibinfo{title}{{Dynamics of the Interplanetary Gas and Magnetic Fields.},} \apj, 128, 664, \dodoi{10.1086/146579}

\bibitem[{I.~G. {Richardson}(2018){Richardson}}]{Richardson2018}
{Richardson}, I.~G. 2018, \bibinfo{title}{{Solar wind stream interaction regions throughout the heliosphere},} \lrsp, 15, 1, \dodoi{10.1007/s41116-017-0011-z}

\bibitem[{I.~G. {Richardson} \& H.~V. {Cane}(2010){Richardson} \& {Cane}}]{2010richardson_RC-list}
{Richardson}, I.~G., \& {Cane}, H.~V. 2010, \bibinfo{title}{{Near-Earth Interplanetary Coronal Mass Ejections During Solar Cycle 23 (1996 - 2009): Catalog and Summary of Properties},} \solphys, 264, 189, \dodoi{10.1007/s11207-010-9568-6}

\bibitem[{O.~M. {Romeo} {et~al.}(2023){Romeo}, {Braga}, {Badman}, {Larson}, {Stevens}, {Huang}, {Phan}, {Rahmati}, {Livi}, {Alnussirat}, {Whittlesey}, {Szabo}, {Klein}, {Niembro-Hernandez}, {Paulson}, {Verniero}, {Lario}, {Raouafi}, {Ervin}, {Kasper}, {Pulupa}, {Bale}, \& {Linton}}]{Romeo2023}
{Romeo}, O.~M., {Braga}, C.~R., {Badman}, S.~T., {et~al.} 2023, \bibinfo{title}{{Near-Sun In Situ and Remote-sensing Observations of a Coronal Mass Ejection and its Effect on the Heliospheric Current Sheet},} \apj, 954, 168, \dodoi{10.3847/1538-4357/ace62e}

\bibitem[{R. {Schwenn}(1990){Schwenn}}]{Schwenn1990}
{Schwenn}, R. 1990, \bibinfo{title}{{Large-Scale Structure of the Interplanetary Medium},} in Physics of the Inner Heliosphere I, ed. R.~{Schwenn} \& E.~{Marsch}, 99

\bibitem[{S.~S. Shapiro \& M.~B. Wilk(1965)Shapiro \& Wilk}]{Shapiro1965}
Shapiro, S.~S., \& Wilk, M.~B. 1965, \bibinfo{title}{An Analysis of Variance Test for Normality (Complete Samples),} Biometrika, 52, 591, \dodoi{10.2307/2333709}

\bibitem[{K.~D.~C. {Simunac} {et~al.}(2012){Simunac}, {Galvin}, {Farrugia}, {Kistler}, {Kucharek}, {Lavraud}, {Liu}, {Luhmann}, {Ogilvie}, {Opitz}, {Popecki}, {Sauvaud}, \& {Wang}}]{Simunac2012}
{Simunac}, K.~D.~C., {Galvin}, A.~B., {Farrugia}, C.~J., {et~al.} 2012, \bibinfo{title}{{The Heliospheric Plasma Sheet Observed in situ by Three Spacecraft over Four Solar Rotations},} \solphys, 281, 423, \dodoi{10.1007/s11207-012-0156-9}

\bibitem[{E.~J. {Smith}(2001){Smith}}]{2001Smith_HCS}
{Smith}, E.~J. 2001, \bibinfo{title}{{The heliospheric current sheet},} \jgr, 106, 15819, \dodoi{10.1029/2000JA000120}

\bibitem[{B.~U.~{\"O}. {Sonnerup} \& M. {Scheible}(1998){Sonnerup} \& {Scheible}}]{Sonnerup1998ISSI_MVA}
{Sonnerup}, B. U.~{\"O}., \& {Scheible}, M. 1998, \bibinfo{title}{{Minimum and Maximum Variance Analysis},} ISSI Scientific Reports Series, 1, 185

\bibitem[{L. {Svalgaard}(1972){Svalgaard}}]{Svalgaard1972_SBlist}
{Svalgaard}, L. 1972, \bibinfo{title}{{Interplanetary magnetic-sector structure, 1926-1971},} \jgr, 77, 4027, \dodoi{10.1029/JA077i022p04027}

\bibitem[{L. {Svalgaard} \& J.~M. {Wilcox}(1976){Svalgaard} \& {Wilcox}}]{Svalgaard1976}
{Svalgaard}, L., \& {Wilcox}, J.~M. 1976, \bibinfo{title}{{Structure of the extended solar magnetic field and the sunspot cycle variation in cosmic ray intensity},} \nat, 262, 766, \dodoi{10.1038/262766a0}

\bibitem[{M. {Temmer}(2021){Temmer}}]{Temmer2021}
{Temmer}, M. 2021, \bibinfo{title}{{Space weather: the solar perspective},} Living Reviews in Solar Physics, 18, 4, \dodoi{10.1007/s41116-021-00030-3}

\bibitem[{M. {Temmer} {et~al.}(2026){Temmer}, {Heinemann}, {Dresing}, {Dumbovic}, \& {Asvestari}}]{Temmer2026}
{Temmer}, M., {Heinemann}, S.~G., {Dresing}, N., {Dumbovic}, M., \& {Asvestari}, E. 2026, \bibinfo{title}{{Coronal Mass Ejection and Heliospheric Current Sheet Interaction Causing a Long-duration Magnetic Field Sector Transition},} \apj, 1004, 203, \dodoi{10.3847/1538-4357/ae6dab}

\bibitem[{M. {Temmer} {et~al.}(2025){Temmer}, {Dumbovi{\'c}}, {Martini{\'c}}, {Cappello}, {Remeshan}, {Matkovi{\'c}}, {Milo{\v{s}}i{\'c}}, {Koller}, {{\v{C}}alogovi{\'c}}, {Susino}, \& {Romoli}}]{Temmer2025}
{Temmer}, M., {Dumbovi{\'c}}, M., {Martini{\'c}}, K., {et~al.} 2025, \bibinfo{title}{{Comparative analysis of two episodes of strongly geoeffective coronal mass ejection events in November and December 2023},} \aap, 695, A58, \dodoi{10.1051/0004-6361/202451479}

\bibitem[{J. {Wang} {et~al.}(2020){Wang}, {Hoeksema}, \& {Liu}}]{wang2020deflection}
{Wang}, J., {Hoeksema}, J.~T., \& {Liu}, S. 2020, \bibinfo{title}{{The Deflection of Coronal Mass Ejections by the Ambient Coronal Magnetic Field Configuration},} Journal of Geophysical Research (Space Physics), 125, e27530, \dodoi{10.1029/2019JA027530}

\bibitem[{J.~M. {Wilcox} {et~al.}(1980){Wilcox}, {Hoeksema}, \& {Scherrer}}]{Wilcox1980}
{Wilcox}, J.~M., {Hoeksema}, J.~T., \& {Scherrer}, P.~H. 1980, \bibinfo{title}{{Origin of the warped heliospheric current sheet},} Science, 209, 603, \dodoi{10.1126/science.209.4456.603}

\bibitem[{R.~M. {Winslow} {et~al.}(2016){Winslow}, {Lugaz}, {Schwadron}, {Farrugia}, {Yu}, {Raines}, {Mays}, {Galvin}, \& {Zurbuchen}}]{Winslowetal2016}
{Winslow}, R.~M., {Lugaz}, N., {Schwadron}, N.~A., {et~al.} 2016, \bibinfo{title}{{Longitudinal conjunction between MESSENGER and STEREO A: Development of ICME complexity through stream interactions},} Journal of Geophysical Research: Space Physics, 121, 6092, \dodoi{10.1002/2015JA022307}

\bibitem[{D. {Winterhalter} {et~al.}(1994){Winterhalter}, {Smith}, {Burton}, {Murphy}, \& {McComas}}]{Winterhalter1994}
{Winterhalter}, D., {Smith}, E.~J., {Burton}, M.~E., {Murphy}, N., \& {McComas}, D.~J. 1994, \bibinfo{title}{{The heliospheric plasma sheet},} \jgr, 99, 6667, \dodoi{10.1029/93JA03481}

\bibitem[{C.-C. {Wu} {et~al.}(2019){Wu}, {Liou}, \& {Lepping}}]{Wu2019}
{Wu}, C.-C., {Liou}, K., \& {Lepping}, R.~P. 2019, \bibinfo{title}{{Solar Cycle Variation of the Heliospheric Plasma Sheet Thickness},} \solphys, 294, 90, \dodoi{10.1007/s11207-019-1464-0}

\end{thebibliography}
\bibliographystyle{aasjournalv7}



\end{document}